\documentclass{llncs}
\usepackage{booktabs}

\usepackage{graphicx}
\usepackage{cite}
\begin{document}
\title{It was never about the language: \newline
       paradigm impact on software design decisions}
\titlerunning{Paradigm impact on design decisions}  
\author{Laura M. Castro\inst{1}}
\authorrunning{Laura M. Castro} 
\institute{Universidade da Coruña\\
           Centro de investigación en TIC (CITIC)\\
           A Coruña, Spain\\
           \email{lcastro@udc.es}}

\maketitle              

\begin{abstract}
Programming languages development has intensified in recent years. New ones are created; new features, often cross-paradigm, are featured in old ones.

This new programming landscape makes language selection a more complex decision, both from the companies points of view (technical, recruiting) and from the developers point of view (career development).

In this paper, however, we argue that programming languages have a secondary role in software development design decisions. We illustrate, based on a practical example, how the main influencer are higher-level traits: those traditionally assigned with programming paradigms.

Following this renovated perspective, concerns about language choice are shifted for all parties. Beyond particular syntax, grammar, execution model or code organization, the main consequence of the predominance of one paradigm or another in the mind of the developer is the way solutions are designed.
\end{abstract}
\section{Introduction}


The current socio-economic context is increasingly considering programming as a key competence in children education~\cite{childrenprogramming}, and predicting a high demand of professionals with programming skills in the short term~\cite{programmersdeficit}. However, no matter the age, the previous experience, or the formal/informal approach, any person who wants to learn to code is immediately confronted with the question of which programming language to choose.

A similar uncertainty is present in IT and software development companies, which struggle in risk evaluation and cost-of-opportunity assessments with regard to sticking to the languages they know and master versus (early) adopting or embracing new languages and technologies. And last but not least, it affects practitioners as well at a personal level, in terms of career decisions for advance and improvement.

In the last decade, a handful of languages have been created that can already be considered mainstream, like Swift~\cite{swift}, Elixir~\cite{elixir}, Kotlin~\cite{kotlin} or Rust~\cite{rust}; a remarkable achievement in such a short time~\cite{stackoverflowdevelopersurvey}. Same happened in the 2000s, with (by now) well-established names such as Go~\cite{go}, C\#~\cite{csharp}, F\#~\cite{fsharp}, Scala~\cite{scala}, Clojure~\cite{clojure}, or VB.NET~\cite{vbnet}.

\begin{table}[ht]
\centering
\begin{tabular}[t]{ccc}
\toprule
Language & Year of creation & Main paradigm \\ \midrule
C\# & 2000 & Object-oriented \\
Clojure & 2007 & Functional \\
Elixir & 2011 & Functional \\
F\# & 2005 & Functional \\
Go & 2009 & Functional \\
Kotlin & 2011 & Object-oriented \\
Swift & 2014 & Object-oriented \\
Rust  & 2010 & Imperative \\
Scala & 2004 & Functional \\
VB.NET & 2001 & Object-oriented \\ \bottomrule
\end{tabular}
\vspace*{0.25cm}
\caption{Popular programming languages which are less than 20 years old.}
\label{tab:languages}
\end{table}

While many of the languages in Table~\ref{tab:languages} are described as multi-paradigm, it is remarkable that half of them can be classified as functional languages. Programming paradigms are the most commonly used means of classification for programming languages, based on their features. The main programming paradigms, imperative and declarative, have been largely considered antagonistic, and it is commonly acknowledged that changing from one domain to the other takes significant mental energy~\cite{paradigm-change,paradigm-change2}, same as mastering the corresponding set of features.

In this paper, we argue that the actual impact of the paradigm-in-mind goes beyond the execution model, code organization, syntax and grammar (Section~\ref{sec:methods}). We show, by means of a simple yet complete and illustrative example, that the influence of the programming language a developer is more used to or intends to use as target reflects on the actual design of the software in a structural and organizational way, and that this affects non-functional characteristics of the software result, such as stability, robustness, or fault-tolerance. We also compare our position to previous analysis of the impact of paradigm or language choice in software products and their qualities (Section~\ref{sec:discussion}).

Our main contribution is a \textit{new perspective on language selection}, meaningful both at the individual and the organizational level: the actual language is less relevant than its main paradigm. In other words, we show how it is the paradigm underneath that drives the design decisions that developers will make, and consequently the key aspect to consider. This realization can alleviate the decision burden of learning or adopting a new programming language, given that paradigm is preserved. It can also motivate technology change, with the goal of shifting the approach to software design.

\section{Methods}
\label{sec:methods}

For decades, the programming landscape was dominated by imperative languages. The imperative paradigm understands programming as a series of commands, instructions to be given to a computer in a specific order. As a result of the execution of said instructions, the computer state is altered and the desired behaviour is obtained. 

In the early and mid-1990s, the prevalence of imperative programming shifted in favour of object-orientation. However, for software developers, this still meant thinking in terms of commands in certain order (for object-orientation is still a \textit{subtype} of imperative programming): but the instructions and the data over which they operate now where shaped into procedures and data fields, with accessibility restrictions of the first over the later (and the first amongst each other) depending on which  ``object'' they were associated to.

It has only been in the new millennium that the functional paradigm has broken its own entry barrier into industry~\cite{fpinindustry,fpinindustry2,fpinindustry3}, even if it had been around for much longer. Often considered in contrast to imperative programming, the functional paradigm understands programming as the composition of functions which do not alter state, but rather offer a return value (which can itself be another function). This deterministic nature is one of the big differences with imperative procedures, which often have side effects due to state alteration.

The first-class citizenship of functions and the restriction of side effects have given solid ground for the argumentation that the functional paradigm favours programs which are easier to reason about, debug and test~\cite{whyfpmatters}. In the age of parallelism and concurrency, this has been seen as an enormous advantage, and is possibly behind the adoption of ``functional traits'' by languages that identify mainly as imperative or object-oriented~\cite{fp-in-oo,fp-in-oo2}, as well as the current popularity of functional languages~\cite{popularity-of-fp}.


However, the perspective that the impact of paradigm choice restricts to the programming levels is very limited. On the contrary, our argument is that said impact is much broader, extending to the higher design of the software, its very own conception. By impacting the software design, paradigm choice affects, for instance, the number and responsibilities of the components that will integrate the solution, its scalability and fault-tolerance.


\subsection{Practical illustration of how ``paradigm thinking'' impacts software design}

To illustrate this central point, we will use a simple example taken from a real project. Let us consider a college environment, and think specifically of last-year students. At most universities, these students would need to carry out a final degree project before graduating. 

Now, the final degree project assignment may vary greatly from institution to institution, even within the same country or state. It might be the case that the student must come up with their own project, or that a set of project proposals are offered and the students request them, with the assignment being made using some objective criteria like average mark on their student record.

If we were to design a system to automatically assign project proposals to students based on certain criteria, the user story that would drive the system is shown in Table~\ref{tab:user-story}.

\begin{table}[ht]
\centering
\begin{tabular}[t]{p{0.8\textwidth}}
\toprule
AS A student \newline
I WANT TO introduce my preferences of project proposals \newline
SO THAT the system can assign me one \\ 
\bottomrule
\end{tabular}
\vspace*{0.25cm}
\caption{User story of the automatic project assignment system.}
\label{tab:user-story}
\end{table}

\begin{figure}
\centering
\includegraphics[scale=0.3]{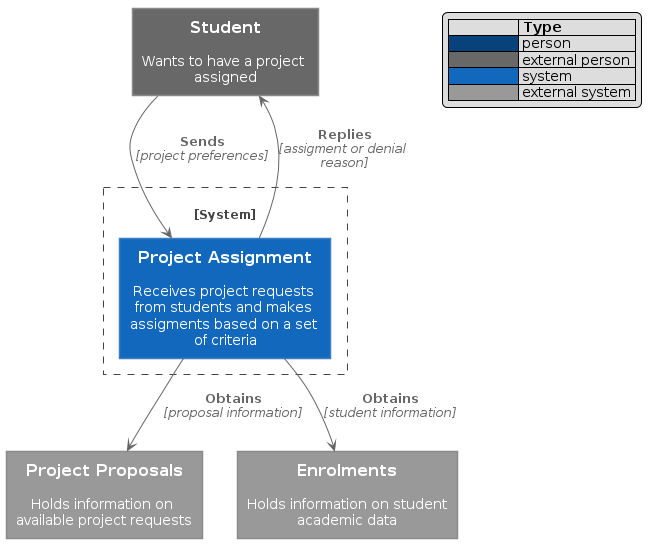}
\caption{Architectural representation of motivation example (C4 context level)}
\label{fig:architecture}
\end{figure}

Let us assume that information like the list of available project proposals is to be retrieved by integration with a different system (possibly, a software used by teachers to create, update and delete said proposals, and also by the corresponding academic committee that would oversee them all), same as the academic information concerning each student (possibly, the enrolment system, which holds not only data on the average marking, but also the concentration that each student is doing, the number of credits the student has passed, etc.), which may play a role as assignment constraints. If so, the overall architecture of the solution will be depicted as shown in Figure~\ref{fig:architecture} in C4 format~\cite{c4}. \\

In the upcoming subsections we analyse the internals of the Project Assignment component to see how paradigm choice affects software design.

\subsubsection{The imperative approach}

A series of commands is nothing else but an algorithm. When developers approach a software problem with an imperative paradigm mindset, they will focus on the algorithm that will solve it. This will reflect in the design of few, very powerful components that:

\begin{itemize}
    \item have unlimited access to all data they need to carry out their task
    \item embed the complete logic of the solution, in a centralised fashion
\end{itemize}

\begin{figure}
\centering
\includegraphics[scale=0.3]{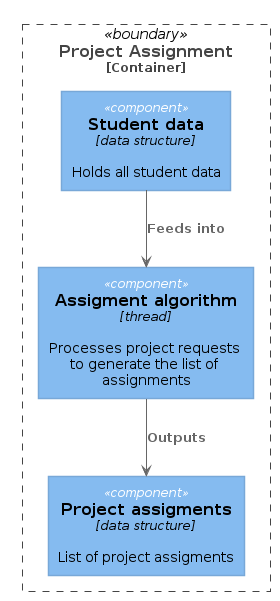}
\caption{Imperative approach design (C4 component level)}
\label{fig:imperative}
\end{figure}

An example of imperative design for the project assignment example is shown in Figure~\ref{fig:imperative}. Aside from the functional aspect, it is worth noting that failure is to be held under the same conditions as the rest of the logic: as part of the algorithm. This means that any fault-tolerance and resilience properties need to be incorporated into the system in one of two ways: either by allowing the system to simply 
``run again'' if something goes wrong (with the consequential waste of time and resources), or to incorporate fault-tolerance management into the problem-solving logic (with the consequential complexity increase that this lack of separation of concerns brings~\cite{concernseparation}).

\subsubsection{The object-oriented approach}

Using objects to encapsulate data and procedures will reflect in the structure of the design by favouring the appearance of more components (objects), each of which is responsible for its own data, both in terms of ownership and in terms of operating with it.

However, it is not so straightforward to know how to distribute responsibilities with regard to business logic as it is to do so with regard to data and relatively small tasks on that data. Even more when object-orientation, for quite some time, did not come hand-in-hand with asynchrony, rather the opposite~\cite{synchronousOO}. Some have even used the term \textit{agent} to differentiate it from the classical \textit{object} to reflect this~\cite{agentvsobject} (cf.~``\textit{agent-oriented} programming''~\cite{aop}).

\begin{figure}
\centering
\includegraphics[scale=0.33]{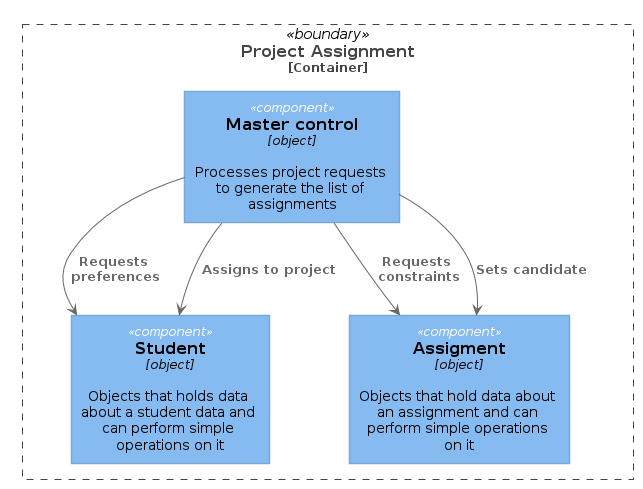}
\caption{Object-oriented approach design (C4 component level)}
\label{fig:oo}
\end{figure}

An example of object-oriented design for the project assignment example is shown in Figure~\ref{fig:oo}. The structure is not as \textit{linear} as in the imperative approach (cf.~Figure~\ref{fig:imperative}): a main orchestrator (control) component will implement a higher-level version of the algorithm, in which object-specific (i.e.~student, assignment) logic is delegated\footnote{In this particular case, these two abstractions will likely hide from the master control the particulars of interacting with the corresponding external systems featured in Figure~\ref{fig:architecture}.}. Similarly, error handling with take place in two levels: internal to the objects, taken care by the objects themselves, and at the algorithm level, which again will be, if present, mixed with the functional logic.

\subsubsection{The functional approach}

If used to capture business logic in the shape of composable functions, a functional developer will approach our project assignment example in a radically different manner. The focus is shifted from data (students, assignments) to processes (proposals, requests).

This will drive, instead of a sequential approximation that goes over the data in order to make a decision, an iterative approximation where partial solutions are offered until a stable one (i.e. a consistent output) is reached.

\begin{figure}
\centering
\includegraphics[scale=0.33]{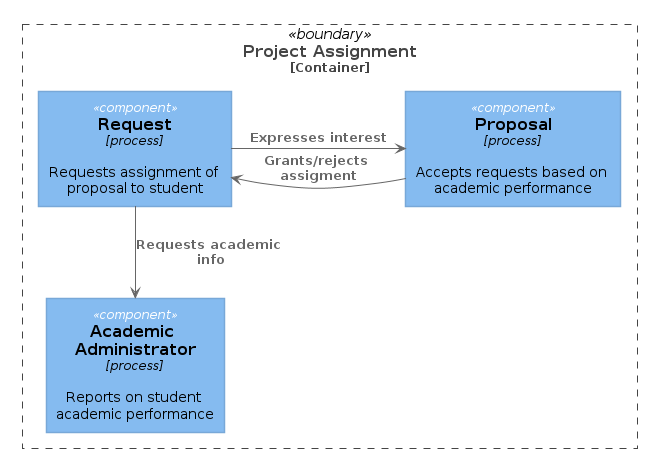}
\caption{Functional approach design (C4 component level)}
\label{fig:functional}
\end{figure}

An example of this functional design is depicted in Figure~\ref{fig:functional}, where the iteration loop that replaces the centralised control of the two previous approaches is shown.

Additional advantages of this solution include the ability of reaching a partial solution in the presence of errors, without explicitly coding error management logic that intertwines with the business logic. If a request or proposal are invalid or become unavailable or corrupt, only the calculation (i.e.~function call) that concerns them will be affected. But given that there is no central control, and that computations are independent and side-effect free, the rest of the execution will still take place.

\begin{figure}
\centering
\includegraphics[scale=0.33]{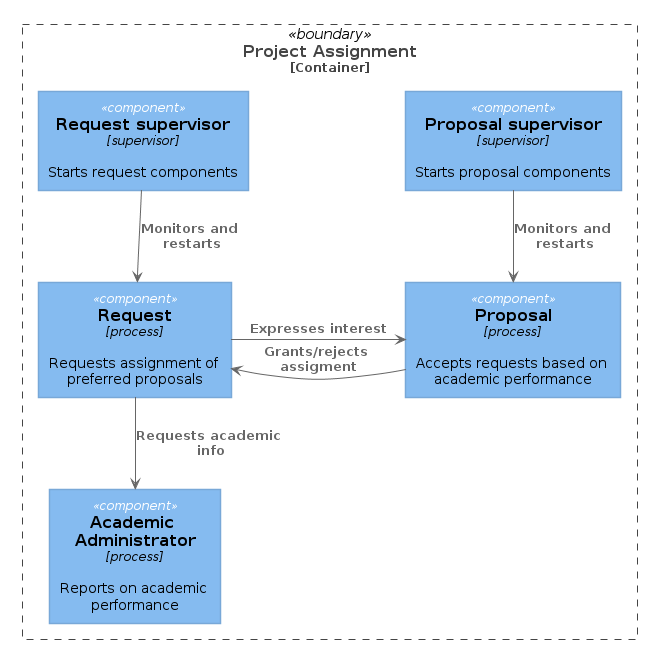}
\caption{Functional approach with supervision (C4 component level)}
\label{fig:functionalsupervised}
\end{figure}

Some functional technologies have taken advantage of this to incorporate powerful fault-tolerance mechanisms that do not interfere with business logic, such as supervision. An enhanced version of the diagram in Figure~\ref{fig:functional} is presented in Figure~\ref{fig:functionalsupervised}, where supervisors transparently provide the ability to retry failed computations.

Moreover, independence of computations and freedom from side-effects also means that operations may take place in different orderings, and even locations, making concurrency and distribution more straightforward, since the sort of data or flow dependencies that typically make them difficult~\cite{concurrencyishard} are not present by design. In our project assignment example, given that the decision making is based not on a temporal criteria (which requests are made first) but on the basis of quantifiable data (i.e.~input information), the order will effectively not affect the result. This, together with the absence of a centralized control, would mean we could approach domains that we do not fully understand, and iterate towards a solution in an incremental manner, by refining the behaviour of simpler functions, rather than a single, large and complex algorithm.

\begin{figure}
\centering
\includegraphics[scale=0.33]{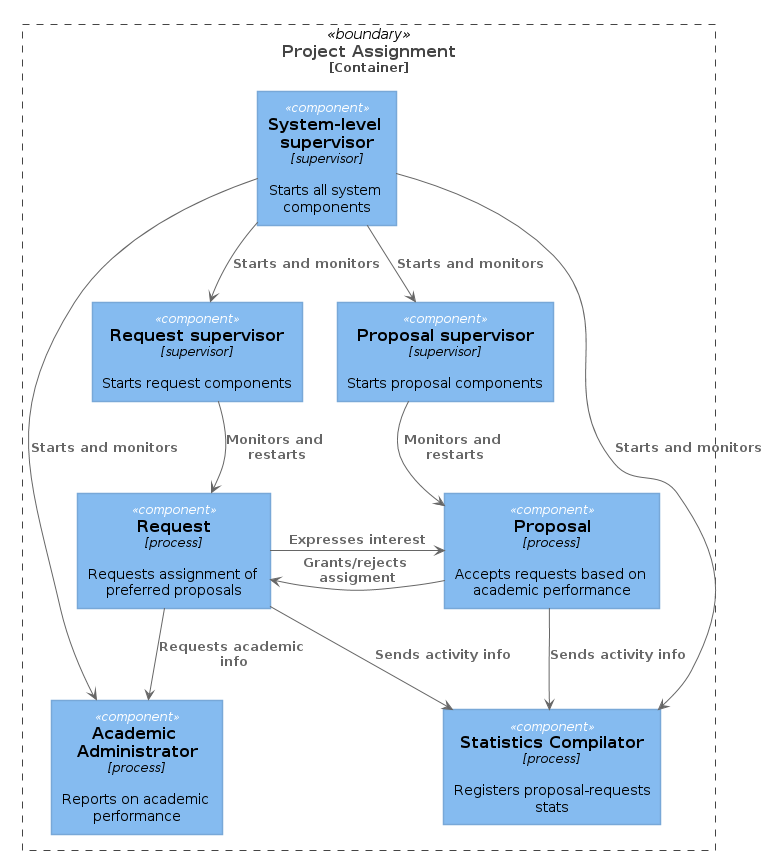}
\caption{Functional approach with explainability (C4 component level)}
\label{fig:functionalexplainable}
\end{figure}

Last but not least, the absence of a single, main, sequential algorithm must not mean the absence of clear and transparent explainability~\cite{explainability}. Once a stable situation is reached (i.e.~constant outputs that show no further changes), the system should have a means to show how that situation was reached, both for demonstrability and for auditing purposes. Figure~\ref{fig:functionalexplainable} shows a last version of the functional approach that embodies such logging for accountability purposes.

\section{Discussion} 
\label{sec:discussion}


That programming languages have an effect on software development has been discussed, both in academia and in more informal forums, for decades~\cite{paradigm-change2}.
However, we can argue that the focus of the debate has been on the internal aspects, both of the developers~\cite{students} (e.g.~the skill set to acquire) and of the code itself~\cite{whyfpmatters,maintainability} (e.g.~its legibility, maintainability, testability, etc.), but not so much on the external aspects of the software that is created (e.g.~its architecture), or even the effect on developers minds and way of thinking and approaching problems.

At a moment in time where soft skills are getting more and more attention~\cite{softskills}, the problem-solving capabilities that software professionals have become much broader than dominating a particular programming language. Similarly, a company's competitive advantage goes beyond collaborative and organizational tools~\cite{tools}. In this context, it is very relevant to ask which one is the key influencer: the programming language or the paradigm?

Programming language popularity has been shown to be hardly related to its internal characteristics~\cite{adoption,future}, rather its application area or business environment~\cite{popularity}.  Also, the programming habits and thought-shaping that programming paradigms can have, have been analysed in the context of paradigm change~\cite{changingparadigms}, but not so much for the possible benefits of maintaining their guidelines regardless of the particular implementation (i.e.~language).

\section{Conclusions}

In this paper, we have argued, in a previously unexplored dimension, that is not the programming language that is primarily relevant in terms of software development, but the paradigm. We have used a simple yet realistic example how this can be the case, but not at code-level, but at much higher abstraction level: that of the software architecture.

We expect that these reflections will open new perspectives, both individual and collective, when it comes to language adoption and technology change.

Of course, the preliminary insights presented in this paper could and should be explored in both analytical and empirical ways, either via developer surveys or analysing the combination of architectural patterns and programming paradigms of open source projects. We intend to continue this line of research in the short term.

\bibliographystyle{splncs03}
\bibliography{biblio}

\end{document}